\documentclass[useAMS,usenatbib]{mnras}

\usepackage{aas_macros}
\usepackage{times}
\usepackage{graphicx}
\usepackage{standalone}
\usepackage{subfig}
\usepackage{multirow}
\usepackage{array}

\newcommand{\xmm}{{\em XMM-Newton}}
\newcommand{\chan}{{\em Chandra}}
\def \psr{PSR\, J2043+2740}

\title[LBT observations of \psr]{Large Binocular Telescope observations of \psr \thanks{The LBT is an international collaboration among institutions in the United States, Italy and Germany. LBT Corporation partners are: The University of Arizona on behalf of the Arizona university system; Istituto Nazionale di Astrofisica, Italy; LBT Beteiligungsgesellschaft, Germany, representing the Max-Planck Society, the Astrophysical Institute Potsdam, and Heidelberg University; The Ohio State University, and The Research Corporation, on behalf of The University of Notre Dame, University of Minnesota and University of Virginia. }}
\author[V. Testa, et al. ]
{\parbox{\textwidth}{V. Testa$^{1}$\thanks{E-mail: vincenzo.testa@oa-roma.inaf.it}, 
R. P. Mignani $^{2,3}$,
N. Rea$^{4,5}$,
M. Marelli$^{2}$,
D. Salvetti$^{2}$,
A. A. Breeveld$^{6}$,
F. Cusano$^{7}$,
R. Carini$^{1}$
} 
\\ \\
$^{1}$ INAF - Osservatorio Astronomico di Roma, via Frascati 33, 00078, Monte Porzio Catone, Rome, Italy \\
$^{2}$ INAF - Istituto di Astrofisica Spaziale e Fisica Cosmica Milano, via E. Bassini 15, 20133, Milano, Italy\\
$^{3}$ Janusz Gil Institute of Astronomy, University of Zielona G\'ora, Lubuska 2, 65-265, Zielona G\'ora, Poland \\
$^{4}$ Institute of Space Sciences (IEECÐCSIC), Carrer de Can Magrans s/n, E-08193 Barcelona, Spain \\
$^{5}$ Anton Pannekoek Institute for Astronomy, University of Amsterdam, Postbus 94249, NL-1090-GE Amsterdam, The Netherlands \\
$^{6}$ Mullard Space Science Laboratory, University College London, Holmbury St. Mary, Dorking, Surrey, RH5 6NT, UK \\
$^{7}$ INAF - Osservatorio Astronomico di Bologna, via Gobetti 93/3, 40129 Bologna, Italy
}

\begin{document}

\date{Accepted 2017 September 24. Received 2017 September 8; in original form 2017 June 30}

\pagerange{\pageref{firstpage}--\pageref{lastpage}} \pubyear{2002}

\maketitle

\label{firstpage}

\begin{abstract}
We present the results of deep optical imaging of the radio/$\gamma$-ray pulsar \psr, obtained with the Large Binocular Telescope (LBT). 
With a characteristic age of 1.2 Myr, \psr\ is one of the oldest (non recycled) pulsars detected in $\gamma$-rays, although with still a quite high rotational energy reservoir ($\dot{E}_{\rm rot} = 5.6 \times 10^{34}$ erg s$^{-1}$). The presumably close distance (a few hundred pc), suggested by the hydrogen column density ($N_{\rm H} \la 3.6 \times 10^{20}$ cm$^{-2}$), would make it a viable target for deep optical observations, never attempted until now. We observed the pulsar 
with the Large Binocular Camera of the LBT. The only object (V=25.44$\pm$0.05)  detected within $\sim$3\arcsec\ from the pulsar radio coordinates is unrelated to it. 
\psr\  is, thus, undetected down to V$\sim$26.6 ($3\sigma$), the deepest limit  on its optical emission. We discuss the implications of this result on the pulsar emission properties. 
\end{abstract}


\section{Introduction}

The launch of the {\em Fermi} Gamma-ray Space Telescope has spurred on the search for pulsars in $\gamma$-rays  
(Grenier \& Harding 2015), yielding over 200\footnote{{\texttt https://confluence.slac.stanford.edu/display/GLAMCOG/}} detections and
triggering multi-wavelength observations. While pulsars are common targets in the X-rays,
they are very challenging targets in the optical
and very few of them have been identified (see Mignani et al.\ 2016 and references therein).
Here we report on Large Binocular Telescope (LBT) observations of an isolated pulsar, \psr,  
detected by both {\em AGILE}  (Pellizzoni et al.\ 2009) and {\em Fermi} (Abdo et al.\ 2010; Noutsos et al.\ 2011).
It was discovered as a radio pulsar (Ray et al.\ 1996) and later on as an X-ray source by \xmm\ (Becker et al.\ 2004), 
although X-ray pulsations have not yet been found. 
\psr\ is one of the very few non-recycled pulsars older than 1 Myr detected in $\gamma$-rays, with a characteristic 
age $\tau_{\rm c} = 1.2$ Myr, inferred from the values of its spin period P$_{\rm s}$=0.096 s and its derivative 
$\dot{P}_{\rm s}= 1.27 \times 10^{-15}$ s s$^{-1}$ (Ray et al.\ 1996).  This also yields  a rotational energy loss 
rate $\dot{E}_{\rm rot} = 5.6 \times 10^{34}$ erg s$^{-1}$ and a surface dipolar magnetic 
field  $B_{\rm s}= 3.54 \times 10^{11}$ G\footnote{Derived from the magnetic dipole model, 
e.g. Kaspi \& Kramer (2016). }.  Although \psr\ does not have a very large spin-down power compared to 
young ($\sim1$--10 kyr) pulsars ($\sim10^{36}$--$10^{38}$ erg s$^{-1}$), it is still a factor of two larger than 
that of middle aged $\gamma$-ray pulsars ($\sim0.1$--0.5 Myr), such as Geminga, PSR\, B0656+14, and PSR\, B1055$-$52,
all detected in the optical  thanks to their distances $\la$ 500 pc (Abdo et al.\ 2013).  
The distance to \psr\ is uncertain owing to the lack of a radio parallax measurement. The radio dispersion measure 
(DM=21.0$\pm$0.1 pc cm$^{-3}$; Ray et al.\ 1996) gives a distance of 1.8$\pm$0.3 kpc from the NE2001 model 
of the Galactic free electron density  (Cordes \& Lazio 2002). A slightly smaller distance (1.48 kpc) is inferred from the model of Yao et al.\ (2017).
The hydrogen column density towards the pulsar obtained from the X-ray spectral fits ($N_{\rm H} \la 3.6 \times 10^{20}$ cm$^{-2}$; 
Abdo et al.\ 2013) suggests a distance of a few hundred pc (He et al.\ 2013), although these estimates depend on the model X-ray spectrum.
Such a distance would make \psr\ a viable target for deep optical observations, never carried out until now,
and might be compatible with the debated association (Noutsos et al.\ 2011) with the Cygnus Loop supernova remnant 
(SNR) at $540^{+100}_{-80}$ pc (Blair et al.\ 2005). 

The structure of this manuscript is as follows: observations and data reduction are described in Sectn.\ 2, whereas 
the results are presented and discussed in Sectn.\ 3 and 4, respectively.

\section{Observations and data analysis}

The \psr\ observations were carried out on  July 5th, 2016 with the LBT at the Mount Graham International Observatory (Arizona, USA) and the 
Large Binocular Camera (LBC; Giallongo et al.\ 2008). The Camera's field of view is 23\arcmin$\times$25\arcmin, with a pixel scale of  0\farcs2255. 
The images were taken  through the filters SDT-Uspec, V-BESSEL, and i-SLOAN, closely matching  the Sloan filters u and i (Fukugita et al.\ 1996), and the Johnson V filter.  
For each filter, three sets of exposures were acquired with exposure times of 20s, 60s and 120s, for a total integration of 5887 s (Uspec and i-SLOAN) and 5376 s (V-BESSEL).  Sky conditions were non-photometric owing to the presence of cirri and the average seeing was around 1\farcs2.
The target was observed with an average airmass around 1.01 and 1.09, and with a lunar illumination of $\sim 1\%$.
Images were reduced with the LBC data reduction pipeline, correcting raw science frames for bias, dark and 
flat fields. A further low-order flat-field correction was obtained from the night sky flats to remove large-scale effects. We, then, corrected the images  for geometrical distortions, 
applying a linear pixel scale resampling. Finally, we stacked all images taken with the same filter 
and used the  master frames to compute the astrometric solution ($\sim$ 0\farcs1 rms).

Since the night was non-photometric, we performed the photometric calibration directly on the science frames by matching stars in public source catalogues. 
In particular, for the SDT-Uspec filter we used a source list  extracted from the \xmm\ Serendipitous Ultra-violet Source Survey Catalogue version 3.0 
(XMM-SUSS3\footnote{{\texttt https://www.cosmos.esa.int/web/xmm-newton/xsa}}), built from observations with the \xmm\ Optical Monitor (OM; Mason et al.\ 2001), whereas for both the V-BESSEL and i-SLOAN filters we used  a source list 
from the American Association of Variable Stars Observers (AAVSO) Photometric All-Sky Survey\footnote{{\texttt http://www.aavso.org/apass}} (APASS).  
All magnitudes are in the AB system (Oke 1974). For all filters, we computed object photometry with the {\textsc DAOPHOT II} software package (Stetson 1994)  
following a standard procedure for source detection, modelling of the image point spread function (PSF), and multi-band source catalogue generation  
(see, e.g., Testa et al.\ 2015). 
After accounting for photometric errors, the fit residuals turned out to be $\sim$ 0.01 magnitudes in all filters, to which we must add the average absolute photometry accuracy of  SUSS3 and APASS, which is $\sim$ 0.05 magnitudes.

\section{Results}

\begin{figure}
\centering
{\includegraphics[width=7cm]{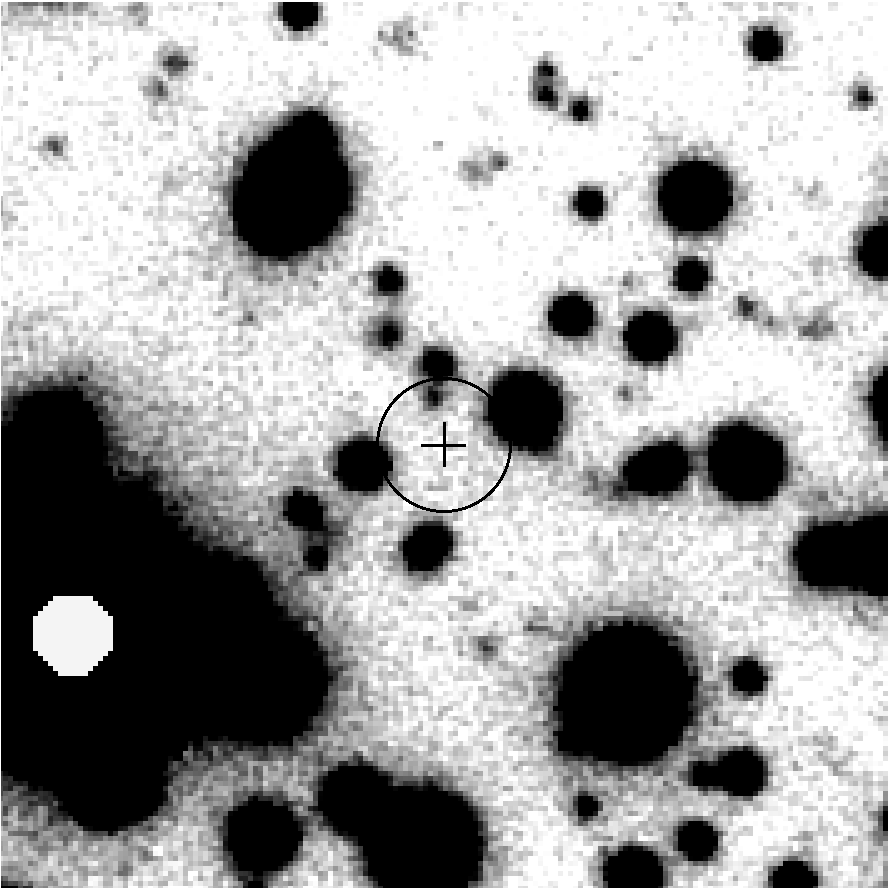}} 
\caption{\label{fc} 
Zoom of the \psr\ field (20\arcsec $\times$20\arcsec) obtained in the i band (5887 s) and  centred on the pulsar position (cross). North to the top, east to the left. 
The length of the cross arms (1\arcsec) is equal to the nominal uncertainty on the radio coordinates. The circle (3\arcsec\ radius) indicates the search region assumed to account 
for the unknown pulsar proper motion.
}
\end{figure}

Fig. \ref{fc}  shows a zoom of the  i-band image centred around the pulsar position. The J2000 coordinates of \psr\ 
used by Noutsos et al.\ (2011) are: $\alpha =20^{\rm h}  43^{\rm m} 43\fs52$; $\delta  = +27^\circ 40\arcmin 56\farcs06$ but with no quoted error. 
The ATNF pulsar catalogue (Manchester et al.\ 2005) reports the same coordinates, with an uncertainty of 0\fs1 and 1\arcsec\ in right ascension and declination, respectively, 
at a reference epoch MJD=49773. 
Owing to timing noise, no updated pulsar coordinates could be computed using the timing model of Kerr et al.\ (2015).  \psr\ has not been observed by \chan, so that we cannot rely on an accurate, model-independent position.
No proper motion has been measured for \psr. Therefore, to account for its unknown angular displacement between the epoch of the reference radio position and that of our LBT observations (MJD=57574), we looked for candidate counterparts within a conservative search region of 3\arcsec\ radius. This is three times as large as the formal radio position uncertainty 
 and roughly equal to the angular displacement expected for a pulsar moving with an average transverse velocity of 400 km s$^{-1}$ (Hobbs et al.\ 2005) at a distance as close as the Cygnus Loop SNR ($540^{+100}_{-80}$ pc; Blair et al.\ 2005).

\begin{figure}
\centering
{\includegraphics[width=8cm]{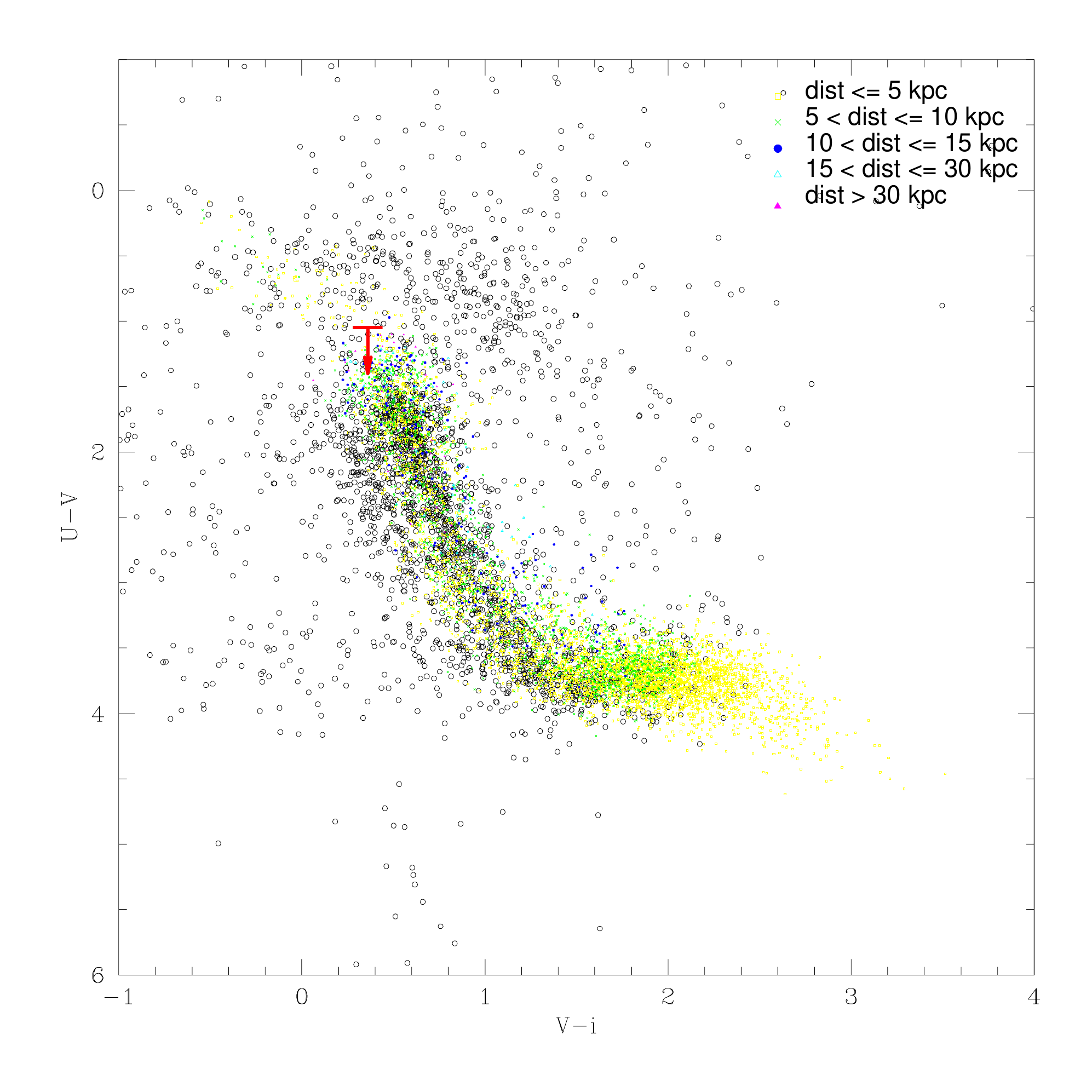}} 
\caption{\label{besancon} Synthetic CCD  for the Galactic stellar population (in colour for different distances) in the pulsar direction simulated with the Besan\c{c}on model (Robin et al.\ 2004), superimposed to the observed one (black). 
The spread in the synthetic CCD is the effect of the distance-dependent reddening correction, applied by the model, and the simulated photometric error.
}
\end{figure}

Only one object is detected within the search region (3\arcsec\ radius) defined above
(Fig. \ref{fc}).  The object is barely visible in the V band and not in the U band, whereas it is clearly detected in the i band. 
Its magnitudes have been computed following the same procedure as described in Sectn. 2 and are V=25.44$\pm$0.05, i=25.08$\pm$0.08, U$>$26.5 (AB).
To investigate the characteristics of the object, we built a U$-$V vs V$-$i colour-colour diagram (CCD) of all objects within 5\arcmin\ from the pulsar position and compared its colours  with respect to the main sequence
(Fig. \ref{besancon}).
Since the field stars are, presumably, at different distances with respect to  the pulsar, the diagram is uncorrected for the reddening. 
The object's colours are V-i = 0.36 $\pm$ 0.09, U-V $>$ 1.06 and are close to those of the main sequence.
This means that it does not stand out for having peculiar colours,  as one would expect for a pulsar, which is usually  characterised by blue colours
(e.g., Mignani \& Caraveo 2001; Mignani et al.\ 2010).
We compared the observed CCD to a synthetic one computed with  the Besan\c{c}on Model of Stellar Population Synthesis
to simulate the Galactic stellar population within a 5\arcmin\ radius around the direction of \psr.
As shown in Fig.\ref{besancon}, the main sequence of the observed CCD is consistent with the model Galactic stellar population, supporting the conclusion that the object 
is a field star rather than the pulsar.
Estimated $3\sigma$-level limiting magnitudes are 26.5, 26.6, and 26.2 in the U, V and i bands, respectively, which we assume as upper limits on the pulsar fluxes.

\section{Discussion}

Our observations of \psr\ are much deeper than those obtained by Beronya et al.\ (2015) with the BTA  (Bolshoi Teleskop Alt-azimutalnyi) 6m telescope,
which only yielded a  $3\sigma$ limit of $R\sim 21.7$ (Vega) on the pulsar flux. The pulsar is obviously too faint to have been detected in the \xmm\ OM images (Becker et al.\ 2004),  with $3\sigma$  limits of B$\approx$21.5 and U$\approx$20.9 (Vega).

\begin{table}
\begin{center}
\caption{Comparison between the optical properties of \psr\ and the two pulsars closest in age detected in the optical. 
}
\label{psr}
\begin{tabular}{llll} \hline
         & B1055$-$52 & J2043+2740 & B1929+10     \\  \hline
$\tau_c$ (Myr)   &  0.5 & 1.2 & 3  \\
$\dot{E}_{\rm rot}$ ($10^{34}$ erg s$^{-1}$) & 3.0 & 5.6 & 0.39 \\
$L_{\rm opt}$ ($10^{27}$ erg s$^{-1}$) &  3.74 & $\la 22.0$  & 1.04 \\ 
$\eta_{\rm opt}$  ($10^{-7}$) & 1.26 & $\la 3.93$ & 2.7 \\
$F_{\rm opt}/F_{\gamma}$ ($10^{-6}$) & 0.88 & $\la 7.14$ & $\ga 31.1$ \\
$F_{\rm opt}/F_{\rm X }$ ($10^{-4}$) & 16.9 &  $\la 5.57$ & 3.4 \\
\hline
 \end{tabular}
\end{center}
\end{table}

We checked whether our limits on the pulsar flux could help to prove or disprove the association with the Cygnus Loop SNR.
In general, the non-thermal optical luminosity $L_{\rm opt}$ of rotation-powered pulsars scales with a power of  the rotational 
energy loss rate (see, e.g. Mignani et al.\ 2012) as $L_{\rm opt} \propto \dot{E}_{\rm rot}^{1.70\pm0.03}$ ($1\sigma$ statistical error).
From this relation, we  estimate a luminosity of $\sim 3.16\times 10^{27}$ erg s$^{-1}$ for \psr, corresponding to a 
magnitude $V\sim 26.2$--26.9  at the distance of the Cygnus Loop SNR,
after  accounting for the interstellar reddening $E(B-V)\la 0.06$,  inferred from the $N_{\rm H}$ (Predehl \& Schmitt 1995).  
Therefore, our detection limit (V$\sim 26.6$) does not determine whether the pulsar is at the distance of the Cygnus Loop SNR, and 
their association remains uncertain.  Pushing the limit on the pulsar brightness down to V$\sim$28 would imply a distance larger 
than $\sim$ 1 kpc for the same predicted optical  luminosity and would  disprove this association. 
Given the lack of a counter-evidence, we assume the pulsar DM-based distance (Yao et al.\ 2017) as a reference. 

We compared our constraints on the optical emission of \psr\ with the properties of other  pulsars of comparable age 
identified in the optical (Table \ref{psr}). Among them, \psr\  lies somewhere in between the middle-aged pulsars, with 
the oldest being PSR\, B1055$-$52  ($\tau_{\rm c} \sim 0.5$ Myr), and the oldish ones, such as PSR\, B1929+10 ($\tau_{\rm c} \sim3$ Myr). 
The V-band optical luminosity of \psr\  is $L_{\rm opt} \la 2.2 \times 10^{28} $ d$_{1.48}^{2}$ erg s$^{-1}$, where d$_{1.48}$ 
is its distance in units of 1.48 kpc (Yao et al.\ 2017). If one assumes that the V-band optical emission is entirely 
non-thermal and rotation powered, its emission efficiency would be $\eta_{\rm opt} \la 3.93\times 10^{-7}$ d$_{1.48}^{2}$. 
For comparison, for a distance of 0.35 kpc (Mignani et al.\ 2010), the V-band optical luminosity of PSR\, B1055$-$52 
would be $3.74 \times 10^{27}$ erg s$^{-1}$ and its emission efficiency $1.26\times 10^{-7}$. We note that the optical 
spectrum of PSR\, B1055$-$52 brings the contribution of both non-thermal emission from the magnetosphere and thermal emission 
from the neutron star surface and is the combination of a power-law (PL) and a Rayleigh-Jeans (R-J) (Mignani et al.\ 2010), 
as observed in other middle-aged pulsars. However, the contribution of the R-J in the V band is about an order of magnitude smaller 
than that of the PL, so that its V-band luminosity is essentially non thermal. 
The $\gamma$-ray energy flux above 100 MeV for \psr\ is  $ F_{\gamma}=(1.18\pm0.12) \times 10^{-11}$ erg cm$^{-2}$ s$^{-1}$ 

(Acero et al.\ 2015), whereas its unabsorbed non-thermal X-ray flux (0.3--10 keV) is 
$F_{\rm X}=0.22^{+0.03}_{-0.11} \times 10^{-13}$ erg cm$^{-2}$ s$^{-1}$ (Abdo et al.\ 2013), which gives an 
 optical--to--$\gamma$-ray flux ratio $F_{\rm opt}/F_{\gamma} \la 7.14 \times 10^{-6}$ and an  optical--to--X-ray flux 
ratio  $F_{\rm opt}/F_{\rm X} \la 5.57 \times 10^{-4}$, where the optical flux $F_{\rm opt}$ has been corrected for the extinction. 
For  PSR\, B1055$-$52,  $F_{\gamma}=(2.90\pm0.03) \times 10^{-10}$ erg cm$^{-2}$ s$^{-1}$  and  
 $F_{\rm X}=1.51^{+0.02}_{-0.13} \times 10^{-13}$ erg cm$^{-2}$ s$^{-1}$ (0.3--10 keV), yielding 
$F_{\rm opt}/F_{\gamma} \sim 0.88 \times 10^{-6}$ and $F_{\rm opt}/F_{\rm X} \sim 16.9 \times 10^{-4}$, whereas for PSR\, B1929+10 
the V-band optical luminosity would be  $1.04 \times 10^{27}$ erg s$^{-1}$ for a 0.31 kpc distance (Verbiest et al.\ 2012) and 
its emission efficiency  $\sim 2.7 \times 10^{-7}$. However, PSR\, B1929+10 has not been observed in the optical but in the near-UV (Mignani et al.\ 2002) where
the spectrum is modelled by a PL with spectral index $\alpha \sim 0.5$. Therefore, its extrapolation  to the optical gives uncertain predictions on the 
unabsorbed V-band flux, and it is not possible to determine whether it decouples into a PL plus a R-J, like in PSR\, B1055$-$52 (Mignani et al.\ 2010).  In this case, 
both the non-thermal optical luminosity and emission efficiency would be overestimated. 
PSR\, B1929+10 has been detected in the X-rays with an unabsorbed non-thermal 0.3--10 keV flux $F_{\rm X}=2.64^{+0.12}_{-0.16} \times 10^{-13}$ erg cm$^{-2}$ s$^{-1}$ 
(Becker et al.\ 2006), but not in $\gamma$-rays down to $2.9 \times 10^{-12}$ erg cm$^{-2}$ s$^{-1}$ (Romani et al.\ 2011),  yielding
$F_{\rm opt}/F_{\rm X } \sim 3.4 \times 10^{-4}$ and $F_{\rm opt}/F_{\gamma} \ga 3.11 \times 10^{-5}$. 

\begin{figure}
\centering
{\includegraphics[width=9.0cm]{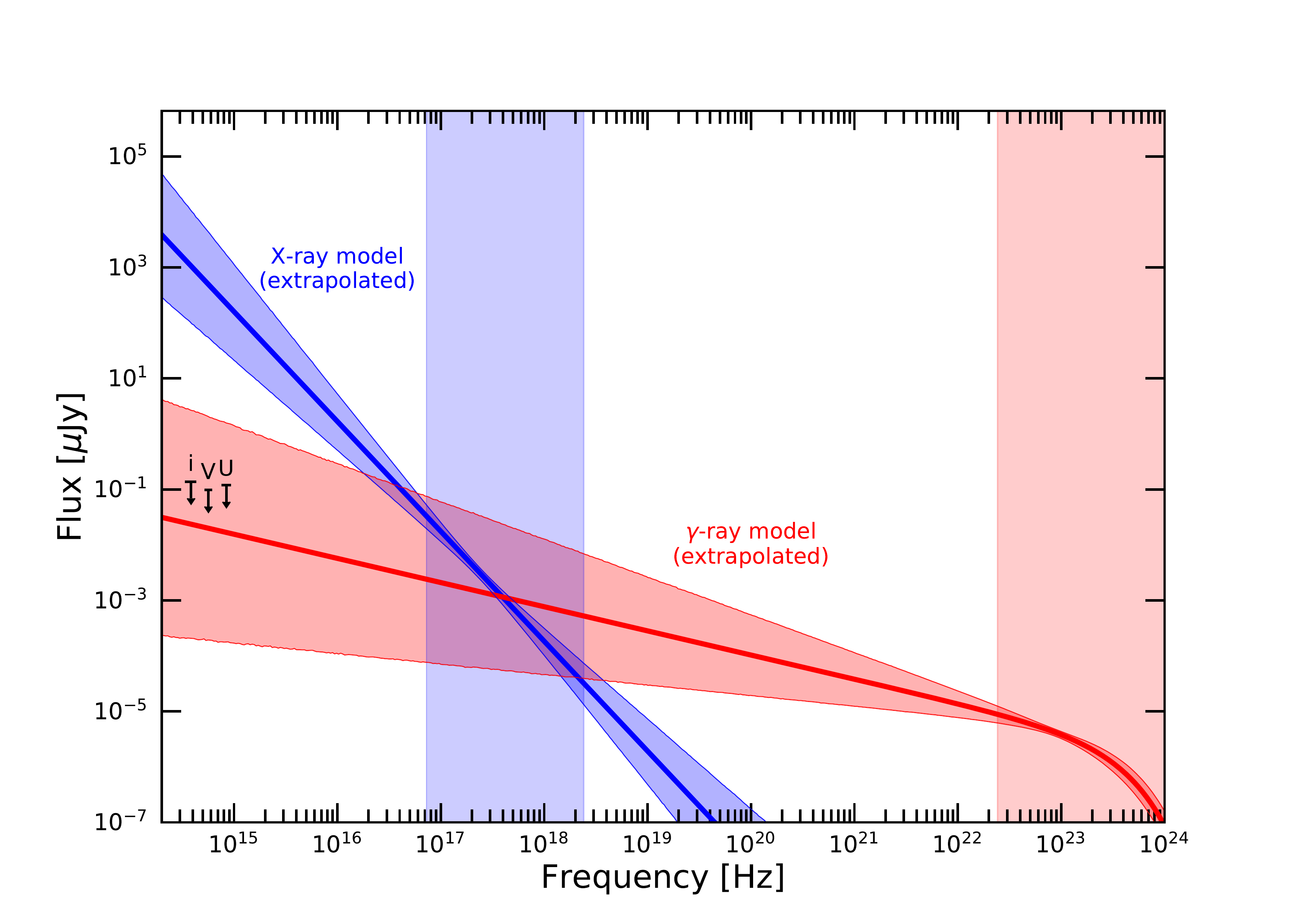}} 
\caption{\label{sed} 
SED of \psr. The extinction-corrected optical flux upper limits are labelled with the band names. The blue and red lines are the extrapolation 
 of the  X and $\gamma$-ray PL spectra, respectively (Abdo et al.\ 2013; Acero et al.\ 2015), with their  $1 \sigma$ errors (dashed lines).  The blue and red regions mark the range where the X and $\gamma$-ray spectra were measured.}
\end{figure}

The $\gamma$-ray spectrum of \psr\ is described by a PL with an exponential cut off, where photon index  $\Gamma_{\gamma}=1.44\pm0.25$ and cut 
off energy $E_{\rm c}= 1.34 \pm0.37$ GeV (Acero et al.\ 2015). The \xmm\ spectrum can be fit by a PL  with $\Gamma_{\rm X} = 2.98^{+0.44}_{-0.29}$
(Abdo et al.\ 2013). The addition of a blackbody component is compatible with the counting statistics, but an $f$-test (Bevington 1969) 
shows no improvement in the fit significance. We compared our optical flux measurements with the extrapolations of the high-energy spectra,
after correcting for  the  reddening using the extinction coefficients of Fitzpatrick (1999).  The spectral energy distribution 
(SED) of \psr\ is shown in Fig. \ref{sed}. As seen in other cases, the extrapolations of the two PL spectra are not compatible with 
each other, implying  a turnover in  the $\gamma$-ray PL at low energies. This is also observed in, e.g.  the 
middle-aged pulsar  PSR\, B1055$-$52 (Mignani et al.\ 2010), although there is no apparent correlation between the presence of a 
turnover and the pulsar characteristic age. 
The optical flux upper limits are below the extrapolation of the assumed X-ray PL spectrum but are not deep enough to rule out that the optical emission 
might  be compatible with the $\gamma$-ray PL extrapolation. This could be a rare case where the $\gamma$-ray and optical spectra are related to each other.

\section*{Acknowledgments}
We thank the anonymous referee for his/her considerate review.
While writing this manuscript, we commemorated the fifth anniversary  of the death of renown Italian astrophysicist Franco Pacini,
 who passed away on January 26th 2012. Franco  authored many seminal publications on neutron stars since right before their discovery  
and was an active promoter of the LBT project. We dedicate our manuscript to his memory. RPM acknowledges financial support from 
an Occhialini Fellowship.  This research was made possible through the use of the AAVSO Photometric All-Sky Survey (APASS), 
funded by the Robert Martin Ayers Sciences Fund.

\label{lastpage}

\end{document}